# A Learned Cost Model-based Cross-engine Optimizer for SQL Workloads

András Strausz
IBM Research
Switzerland
andras.strausz@zurich.ibm.com

Niels Pardon
IBM Research
Switzerland
par@zurich.ibm.com

Ioana Giurgiu
IBM Research
Switzerland
igi@zurich.ibm.com

## ABSTRACT

Lakehouse systems enable the same data to be queried with multiple execution engines. However, selecting the engine best suited to run a SQL query still requires a priori knowledge of the query's computational requirements and an engine's capabilities, a complex and manual task that only becomes more difficult with the emergence of new engines and workloads. In this paper, we address this limitation by proposing a cross-engine optimizer that can automate engine selection for diverse SQL queries through a learned cost model. Optimized with hints, a query plan is used for query cost prediction and routing. Cost prediction is formulated as a multi-task learning problem, and multiple predictor heads, corresponding to different engines and provisionings, are used in the model architecture. This eliminates the need to train engine-specific models and allows the flexible addition of new engines at a minimal fine-tuning cost. Results on various databases and engines show that using a query's optimized logical plan for cost estimation decreases the average Q-error by even 12.6% over using unoptimized plans as input. Moreover, the proposed cross-engine optimizer reduces the total workload runtime by up to 25.2% in a zero-shot setting and 30.4% in a few-shot setting when compared to random routing.

## 1 INTRODUCTION

Data has gone from being scarce to being super-abundant. Never before has it been so easy to collect large data quantities, due to the large-scale infrastructures available in the cloud. However, the increasing workload diversity in modern use-cases (i.e., lately seeking to harness unstructured data to fuel AI innovations) has led to the proliferation of specialized data management systems, each targeted to narrow types of workloads. For example, Postgres excels at executing SELECT queries by using indices, but significantly lags behind Spark for general-purpose batch processing where parallel full scans are key. Presto is built for ad-hoc and interactive workloads, whereas Spark pays the penalty of always having to span and shut down workers as soon as workloads start or finish.

This has led to siloed systems, high maintenance costs and wasted engineering cycles. Even worse, the byproducts of this fragmentation – incompatible APIs, disparate functionality, inconsistent semantics – impact the end users, who commonly need to interact with multiple distinct systems to complete their tasks and to have expert knowledge to use them appropriately.

To alleviate some of these caveats, data management systems have seen a significant shift, from monolithic designs to modular approaches. Recent studies [12, 24] conceptualize a composable data system constructed from multiple independent layers: (1) the user-facing APIs, (2) the optimization layer, (3) the execution layer, and (4) the storage layer. Such cross-platform systems are horizontally extendable, making it easy to include various execution engines or storage systems and to express workloads in different dialects. However, how and where to execute these workloads in a cost/performance-optimal manner remains highly challenging.

**Contributions.** To overcome the above limitation, we propose optimizing engine selection in a lakehouse for SQL workloads through a learned cost model (LCM). The optimizer first applies traditional query-rewriting techniques to supply an optimized logical plan to the LCM, which we show to be beneficial for the downstream tasks of query cost prediction and routing. Cost prediction is formulated as a multi-task learning problem, using a Graph Neural Network (GNN) architecture to compute a general query representation. The resulting embedding is shared among multiple predictor heads corresponding to different engines and their respective provisionings, thus eliminating the need to train engine-specific LCMs.

The optimizer is evaluated in a lakehouse system with five different engine configurations on various synthetic and real-world databases (DBs). In a zero-shot setting, its query-to-engine routing reduces the workload total runtime by up to 25.2% over a random routing. In a few-shot setting, results are even better and the optimizer's routing outperforms random routing by even 30.4%. These improvements translate to tens of minutes saved in execution even for small databases, such as IMDB or TPC-H. Lastly, experiments on introducing a new engine provisioning showcase the optimizer's ability to flexibly add a new predictor head in the LCM at a cheap fine-tuning cost, by training it only on 250 queries.

## 2 BACKGROUND

**Polystores and Federated Data Management Systems.** In line with our objective, polystores [2, 4, 10, 25, 30, 37] and federated DBs [7, 14, 26, 35] also aim to distribute query workloads across heterogeneous engines. Specifically, they optimize queries within a given set of engines and hardware configurations. For example, RHEEM [2] requires additional work to specify selectivity and cost templates for each operator when including a new engine, which can quickly become a burden for extension. In contrast, we enable the easy addition of a new engine to the underlying infrastructure to support the user's workload.

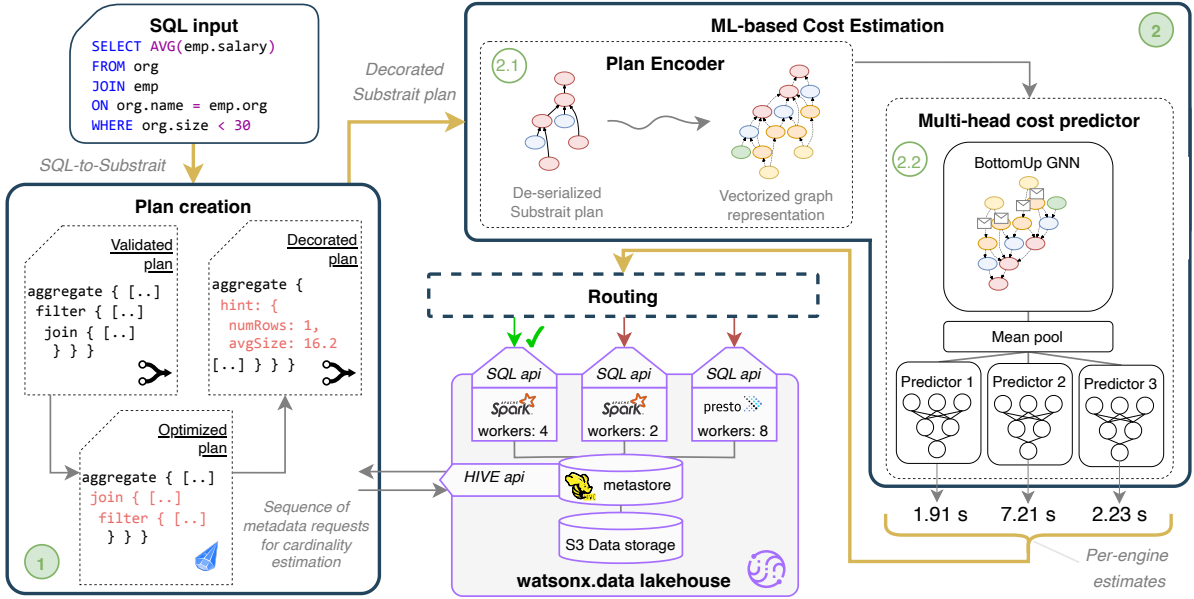

Figure 1: Overview of the cross-engine optimizer.

**LCMs for Query Optimization and Cost Estimation.** Recently, LCMs [5, 8, 15, 20, 21, 23, 32, 38, 39] that aim to learn and enhance the behavior of the DB engine's optimizer have been proposed. Some [5, 20, 23] use the traditional optimizer's hints to improve the optimization procedure. Others [8, 21, 32, 38] attempt to fully replace the query otimizer with a learned query rewriter. Finally, a variety of approaches [13, 31, 33] estimate query cost with LCMs. Stage [31] proposes a hierarchical modeling strategy, where either instance-level or global models are used for the task. For the latter, the Graph Neural Network (GNN) architecture described in [13] is used. Similarly, BRAD [33] uses the same architecture to tackle query-to-engine routing. To select the cost-optimal engine, it considers multiple cost factors, some estimated by closed-form functions, and others by learned models. Most importantly, for execution time prediction, it uses unoptimized logical query plans. Each execution engine is considered separately, so an individual predictor model would need to be trained for each. While we are inspired by the bottom-up GNN architecture, we propose a multi-head predictor that simultaneously predicts execution times for all supported engines and allows new engines to be easily added by fine-tuning a new predictor head on a small volume of queries.

**LCM Architectures.** Various architectures have been explored for LCMs, ranging from flat vectors [11, 16] with Multi Layer Perceptrons (MLP), to Tree Convolutional Networks [22], Recurrent Neural Networks [29, 34] and Transformer models [36]. More recently, GNNs have gained popularity due to their natural representation of query structures. A message passing algorithm over the query execution plan has been proposed in [13]. Similarly to sequence models, by adjusting the message passing order to follow the execution plan's topology, nodes receive information from their subtree. Thus, the computed hidden embedding of a node is a representation of this subtree. Consequently, the root node's embedding can be used as a representation of the complete input graph. Embeddings are computed using node-specific MLPs, and messages are aggregated by summing. Finally, LLMs have been used to embed the query text [3], because of their understanding of predicated or even complete SQL statements. These embeddings can be used either as a complement to other query or predicate embeddings computed using numeric features or even as standalone representations.

## 3 CROSS-ENGINE OPTIMIZER

The cross-engine optimizer acts as a middleware and interacts with the underlying system's engines and metadata provider in a lakehouse. Its objective is, upon receiving a user query, to estimate the query's execution time on each engine using an LCM and then use those for engine selection.

The architecture of the cross-engine optimizer is shown in Figure 1. The input query is first received by the ① *Plan creation* module, which transforms it into an optimized Substrait [28] plan with cardinality hints. This plan is forwarded to the ② *ML-based cost-estimation* module, which encodes the deserialized plan and predicts the query's execution time for each engine. Finally, these predictions are used for *Routing*.

### 3.1 Plan Creation

During plan creation ①, the SQL query is transformed into an optimized Substrait plan with hints containing cost-related information for each relation. The goal is to (1) **reduce variance between plans** and ensure the (2) LCM receives an **accurate representation** of the query's plan and estimated cardinalities.

First, the system verifies the query against the database schema stored in the metastore. Next, it generates the initial Substrait plan, in which the order of operations is solely determined by the SQL text. A plan-trimming step is then applied, ensuring that only referenced fields and tables are included. The trimmed Substrait plan is further converted into a Calcite [6] plan for query optimization.

The optimization procedure first applies predicate pushdown along with traditional techniques to merge relations that can be combined and simplify predicates. Next, a greedy cost-based algorithm is applied to optimize the join order, minimizing the intermediate schema after assigning each join. At any point in the algorithm, the two relational sub-trees estimated to produce the least number of rows are joined. This algorithm is likely to result in "bushy" join trees over left-deep trees, which is preferred to reduce the overall depth of the tree and, consequently, the number of message passing rounds during cost estimation.

Finally, the optimized plan is extended with hints, indicative of the cost of each relational node. These hints are computed following the selectivity estimation rules defined by Selinger's method [27] and are later used during the featurization step. Any data information, such as cardinalities of tables or average column width, is extracted from the lakehouse's metastore.

## 3.2 Cost Estimation with LCM

The cost estimation module ② predicts the execution time of the input query for each engine in the lakehouse. The cost predictor is designed with two requirements: (1) **Database-agnosticity**: it must support prediction across databases with varying schemas; (2) **Support for a variable number of execution engines**: the modeling approach should provide (a) per-engine execution time estimates without requiring the training of engine-specific LCMs, as well as (b) adding a new engine and its corresponding predictor without expensive data generation or complete retraining.

We adopt the Bottom-up GNN algorithm from [13], but use it to learn a general query representation rather than for direct cost prediction. The GNN architecture leverages both node-level feature embeddings and the query plan's graph structure to inform its predictions, making it particularly well-suited for cost prediction. To generate the query embedding, we extend the original Bottom-up algorithm and apply mean pooling over all relation nodes in the plan instead of relying solely on the final node's embedding. This choice is driven by the fact that the information propagation degrades with the depth of the tree, which affects the downstream task. Mean pooling counteracts this by ensuring that information is received from each relation node in the tree.

Multiple predictors can use the same embedding by decoupling the query embedding from downstream prediction tasks. This enables multi-task learning [9], where each task predicts the execution time for a specific engine provisioning. These prediction tasks are related and reinforce one another, helping the shared embedding capture a general representation of the query. The engine-specific predictor heads learn to associate the general embedding with the particular characteristics of each engine for accurate cost estimation. The primary benefit of this approach is that it avoids training separate entire GNNs for every engine. Specifically, when a new engine is added to the system, only its predictor head must be trained while the embedding model remains fixed.

| Node type | Features |
|---|---|
| Table | numRows, avgRowSize |
| Field | numNulls, numDistinctVals, dataType★, avgColSize, maxColSize |
| Literal | dataType★, size, isCasted |
| Operation | operationType★ |
| Relation | numRows, avgRowSize, relationType★ |

**Table 1: Encoding of different node types. The ★ denotes one-hot encoded categorical fields.**

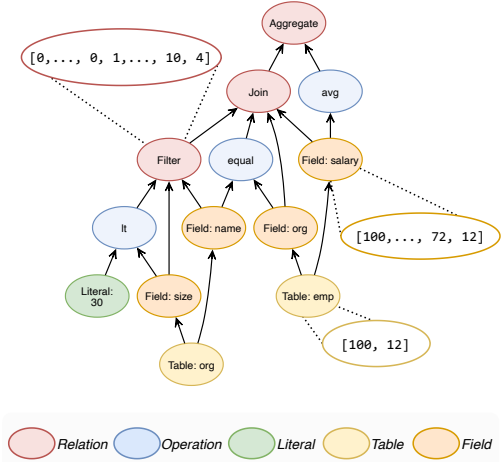

**Figure 2: Encoded graph representation of the query shown in Figure 1.**

*3.2.1 Plan encoding.* The encoding module ②.1 de-serializes the decorated Substrait plan and transforms it to a vectorized graph representation, which serves as the input to the GNN. To this end, we first explicitly expand the Substrait tree to include intermediate schemas during query execution. Finally, each node is encoded to a flat vector containing relevant information about the node's cost.

*Tree structure.* The final representation of the query is a heterogeneous graph containing five different node types: *Relation*, *Operation*, *Literal*, *Field*, and *Table* nodes. Substrait uses implicit schema references in each relation and operation. The implicit schema is manipulated throughout the execution tree encoded by the `emit` field of each relation. To directly include field accesses, a `TableScan` scan operation is expanded to separate *Table* and *Field* nodes. Similarly, literals are included as distinct nodes. Furthermore, Substrait's implicit field references are converted into direct accesses materialized by edges from *Field* to *Relation* nodes.

*Node encoding.* In addition to creating the graph structure, featurization includes converting each node object to a type-specific, fixed-length numeric vector. *Table*, *Field* and *Literal* nodes are solely featurized through attributes about the data they describe.

**Algorithm 1** Multi-task predictor with Bottom-up GNN encoder

1: **Input**: vectorized query graph **x**, set of engines $\mathcal{E}$
2: **Output**: per-engine execution time prediction
3: # Create query graph embedding using Bottom-up GNN
4: **for** $v \in$ input graph **do**
5:     $\mathbf{h}_v \leftarrow \text{EncoderMLP}_T(\mathbf{x}_v)$
6: **for** $v \in$ topological order **do**
7:     $\mathbf{h}'_v \leftarrow \text{HiddenMLP}_T\Big(\sum_{u \in \text{children}(v)} \mathbf{h}'_u \oplus \mathbf{h}_v\Big)$
8: # Mean-pool over embeddings of Relation nodes
9: $\mathbf{h}_{query} \leftarrow \text{mean\_pool}(\{\mathbf{h}'_v : v \in Relation\})$
10: # Prediction using separate predictor heads
11: **for** $i \in 1 \dots |\mathcal{E}|$ **do**
12:     $\hat{\mathbf{y}}[i] \leftarrow \text{PredictorMLP}_i(\mathbf{h}_{query})$
13: **return** $\hat{\mathbf{y}}$

For *Operation* and *Relation* nodes, the featurization also includes the type of the expression (such as Join, Filter for *Relation* or max, +, -, etc. for *Operation*) as a one-hot encoded vector. For *Relation*, each relation included in the Substrait specification is represented as a distinct category. *Operation* expressions are assigned to categories via a predefined static mapping that groups expressions with similar computational complexity. Log-normalization is applied to all continuous features to avoid large differences in scale between features. Table 1 lists the features considered for each node type and Figure 2 depicts an example output of the featurization process.

*3.2.2 LCM architecture.* The multi-head cost predictor ②.② first employs a GNN to convert the encoded query plan into a low-dimensional embedding. This embedding is fed into each predictor head, producing an execution time estimate for the respective execution engine. The embedding heads are implemented as Multi-layer Perceptrons (MLPs). Algorithm 1 provides pseudo code for the multi-task inference process.

The algorithm first projects nodes in the input graph to a common vector space using type-specific encoders. In particular, for each node with type $T \in \{Relation, Operation, Literal, Field, Table\}$, the corresponding encoder $\text{EncoderMLP}_T : \mathbb{R}^{d_T} \to \mathbb{R}^{d'}$ is applied (lines 4-5). Afterwards, in the message passing phase, messages are propagated through the tree in topological order, starting from leaf nodes and progressing towards the last *Relation* node. At each step, nodes send messages to their parent nodes once they have received all messages from their children. After a node receives messages, it updates its hidden embedding by first concatenating it to the sum of received embeddings and feeding this combined embedding through a second, type-specific MLP, $\text{HiddenMLP}_T : \mathbb{R}^{d'+d'} \to \mathbb{R}^{d'}$ (lines 6-7). The mean-pooling operation is then applied over all *Relation* nodes in the graph (line 9) to arrive at the final, learned representation of the query. This representation is then received by the engine-specific predictor heads, $\text{PredictorMLP}_i : \mathbb{R}^{d'} \to \mathbb{R}$ to compute the final estimates.

Let $\mathcal{E} = \{e_1, ..., e_{|\mathcal{E}|}\}$ be a set of engines. For a query **x**, encoded in a vectorized graph format, we record corresponding measurements $\mathbf{y} \in \mathbb{R}^{|\mathcal{E}|}$, where the *i*-th component $\mathbf{y}[i]$ is the execution time of the query measured on engine $e_i$. Furthermore, let $\text{BottomUpGNN}(x) : \mathcal{G} \to \mathbb{R}^{d'}$ represent the complete bottom-up message passing algorithm, including mean-pooling (lines 4-9) and producing a $d'$-dimensional embedding of the input graph. During training, each task (i.e., predicting the execution time for a specific engine configuration) is weighted equally. Namely, for a loss function $\mathcal{L}(\cdot, \cdot)$ between predicted and measured execution time, we compute the prediction error for backpropagation as:

$$l = \frac{1}{|\mathcal{E}|} \sum_{i \in \{1,...,|\mathcal{E}|\}} \mathcal{L}(\text{PredictorMLP}_i(\text{BottomUpGNN}(\mathbf{x})), y[i])$$

|  | Raw Size (GB) | Parquet Size (GB) | #Tables | #Rows (M) | #Rel |
|---|---|---|---|---|---|
| TPC-H | 10 | 3.2 | 8 | 86.6 | 8 |
| TPC-DS | 10 | 4.2 | 24 | 191.5 | 102 |
| IMDB | 3.6 | 1.8 | 23 | 74.3 | 17 |
| Stack Overflow | 4.4 | 2.1 | 9 | 19.3 | 12 |
| Donor | 1.7 | 0.75 | 4 | 7.5 | 4 |

Table 2: Summary of datasets. *#Rel* is the number of foreign-key constraints taken into account for query generation.

During training, an adjusted form of Q-error is employed as the loss function, computing the standard Q-error for positive estimates and assigning an arbitrarily large penalty for negative estimates.

## 4 EVALUATION
### 4.1 Methodology

*4.1.1 Environment.* All experiments have been conducted on a cluster with dual-socket compute nodes, each hosting 2 Intel Xeon E5-2683 v4 CPUs and 768GB of RAM. The cluster is running on OpenShift, where the watsonx.data [1] lakehouse is hosted. Since watsonx.data currently natively supports only PrestoDB and SparkSQL, we evaluate on these 2 engine types with 4 provisionings (1 and 4 worker nodes, respectively). Furthermore, all caching capabilities of PrestoDB are disabled to ensure that measurements are independent. Data is stored in MinIO buckets in parquet format, and a Hive catalog is used to keep and distribute metadata inside the lakehouse. For the experiment introducing a new engine, a PrestoDB provisioning with 8 worker nodes is considered.

*4.1.2 Data Collection.* 5 different datasets were selected for evaluation: TPC-H and TPC-DS with a scale factor of 10, the IMDB dataset from the JOB [17], a one-year data dump of Stack Overflow and the donor dataset from the BIRD-SQL benchmark [18]. Each dataset contains multiple foreign key relationships and column types, ranging from simple numeric and text values to dates and timestamps. Some of the key statistics of the datasets are summarized in Table 2.

Since traditional benchmarks include at most a few hundred queries, insufficient for the LCM training, we use the synthetic query generator from [13]. Following prior works, queries are limited to at most 3 joins and a maximum runtime of 1 minute to allow efficient training data collection. However, the generated queries include predicates on timestamp and date columns as well as subqueries and predicates on aggregates (HAVING statements). As our cross-engine optimizer relies on Calcite's grammar-driven SQL parser and its featurization process covers the full range of relations and data types defined by Substrait, this is the first attempt to support such a broader spectrum of queries systematically.

|        | EncoderMLP$_T$: | (input_dim, 64, 96, 144, 216, 256) × 5 |
| GNN:   | HiddenMLP$_T$:  | (512, 384, 384, 384, 256) × 5 |
|        | PredictorMLP$_e$: | (256, 174, 121, 85, 59, 1) × #engines |
| Set-based: | EncoderMLP$_T$: | (input_dim, 64, 96, 144, 216, 256) × 5 |
|        | PredictorMLP$_e$: | (1024, 174, 121, 85, 59, 1) × #engines |

Table 3: Layer sizes of each MLP used for the GNN and set-based architectures. *input_dim* refers to the encoded dimension (number of features) of each node type.

|        | q$_{med}$ | q$_{mean}$ | q$_{p95}$ |
|--------|-----------|------------|-----------|
| Ours   | 1.21      | 1.47       | 2.43      |
| GNN.UP | 1.24      | 1.55       | 2.71      |
| SB.OP  | 1.21      | 1.47       | 2.45      |

Table 4: Unseen queries: prediction accuracy on unseen queries.

|        | IMDB | | | Stack Overflow | | | TPC-H | | |
|--------|------|------|------|------|------|------|------|------|------|
|        | q$_{med}$ | q$_{mean}$ | q$_{p95}$ | q$_{med}$ | q$_{mean}$ | q$_{p95}$ | q$_{med}$ | q$_{mean}$ | q$_{p95}$ |
| Ours   | 1.51 | 1.81 | 3.57 | 1.44 | 1.95 | 4.23 | 1.40 | 1.71 | 3.16 |
| GNN.UP | 1.67 | 2.07 | 4.36 | 1.45 | 1.95 | 4.35 | 1.49 | 1.84 | 3.52 |
| SB.OP  | 1.53 | 1.90 | 4.13 | 1.47 | 1.97 | 4.26 | 1.43 | 1.80 | 3.32 |

Table 5: Zero-shot: prediction accuracy of zero-shot models.

|        | IMDB | | | Stack Overflow | | | TPC-H | | |
|--------|------|------|------|------|------|------|------|------|------|
|        | q$_{med}$ | q$_{mean}$ | q$_{p95}$ | q$_{med}$ | q$_{mean}$ | q$_{p95}$ | q$_{med}$ | q$_{mean}$ | q$_{p95}$ |
| Ours   | 1.37 | 1.56 | 2.76 | 1.35 | 1.73 | 3.42 | 1.29 | 1.55 | 2.70 |
| GNN.UP | 1.45 | 1.73 | 3.42 | 1.43 | 1.87 | 3.99 | 1.31 | 1.63 | 2.87 |
| SB.OP  | 1.37 | 1.57 | 2.72 | 1.38 | 1.82 | 3.64 | 1.29 | 1.59 | 2.63 |

Table 6: Few-shot: prediction accuracy of few-shot models.

For each dataset, 5000 queries are generated and executed on each execution engine.

*4.1.3 Evaluation Scenarios.* To thoroughly evaluate both the proposed LCM's accuracy in predicting execution time and its effectiveness for routing queries, we consider 4 scenarios:
*(1) Unseen queries*: all 5 datasets are merged for training and testing. For evaluation, 1000 queries are held out from each dataset.
*(2) Zero-shot*: the LCM is evaluated on a dataset excluded from training. We report cross-validation results in which, for each fold, 4 datasets serve as the training set and the fifth is used for evaluation.
*(3) Few-shot*: the LCM is fine-tuned on a small subset of queries drawn from the test dataset. Only the predictor heads are updated, while the shared embedding model remains fixed. We use 250 (∼5%) queries for few-shot experiments, split equally between training and validation.
*(4) New engine*: a new predictor head is trained for the new engine in a few-shot setting. This scenario addresses the case where the current provisioning is insufficient for running the workload under a preferred time or cost budget.

## 4.2 Results on LCM's Accuracy

We begin by analyzing the proposed LCM's estimation accuracy in each scenario, using the Q-error metric, averaged over all considered engines. The metric q$_{med}$ for a set of queries $Q$ and a set of engines $\mathcal{E}$ is computed as:

$$q_{med} = \frac{1}{|\mathcal{E}|} \sum_{i \in 1...|\mathcal{E}|} median\left(\left\{max\left(\frac{pred_q}{true_q}, \frac{true_q}{pred_q}\right) : q \in Q\right\}\right)$$

Results for IMDB, Stack Overflow and TPC-H are presented below. Appendix 6.2 includes the evaluation of the remaining datasets.

*4.2.1 Baselines.* We define 2 baselines to compare against:
*(1) GNN.UP*: the LCM is trained with *validated plans* and no optimizations. Specifically, input plans are produced by applying field trimming to the original Substrait plan and converting subqueries into joins. This baseline is closest to BRAD [33], which creates the input graph directly from the SQL text.
*(2) SB.OP*: we implement an adjusted version of the set-based model proposed by Kipf et al. [16] with optimized plans. Details are provided in Appendix 6.1.

*4.2.2 Results on unseen queries.* Evaluating on queries targeting the same underlying databases as those used for training leads to generally accurate predictions, as reported in Table 4. Note that a median Q-error of 1.21, which is achieved for both our proposed model and for the set-based architecture, means that, on average over each predictor head, 50% of the estimates deviate by no more than 21% from the true, measured execution time.

Comparing optimized and unoptimized plans with the GNN architecture, we see a 5.2% relative reduction in q$_{mean}$ and a 10.3% reduction in the tail error q$_{p95}$ in favor of optimized plans. These results reinforce the idea that the input plans for the LCM can be improved through traditional query optimization techniques. The differences between model architectures are minor and only present in the tail error by a 0.2 difference in q$_{p95}$. Thus, in this scenario, the GNN architecture cannot extract a considerably larger amount of additional information from the plan's structure compared to the simpler set-based model. Furthermore, the accuracy of different predictor heads shows only low variance. This implies that the learned embedding, whether created by the GNN or the set-based model, is general enough to be used for predicting execution times on different engines.

*4.2.3 Results for zero-shot setting.* Results are reported in Table 5. Zero-shot models exhibit substantially higher Q-errors than models trained and evaluated on the combined dataset. Nevertheless, similar trends are observed in the zero-shot setting, where query optimization also yields more accurate predictions compared with those obtained using unoptimized plans. In particular, using optimized plans with the GNN architecture improves on the average Q-error by 9.6% and by 18.2% in q$_{p95}$ for the IMDB dataset over using unoptimized plans. On the other hand, results for the Stack Overflow dataset show a difference only in the tail error. For this dataset, all considered settings lead to similar metrics, and generally high tail error. We also observe that the GNN architecture generalizes better than the set-based model, achieving superior results on all but one dataset.

*4.2.4 Results on few-shot setting.* The few-shot setting is designed to enable the predictor heads to capture dataset-specific cost characteristics by fine-tuning and thus produce more accurate predictions. Comparing the few-shot results (Table 6) with the zero-shot results shows that fine-tuning indeed improves the LCM's accuracy, as reflected by improvements in each metric across all configurations.

|       | **IMDB** | | | **Stack Overflow** | | | **TPC-H** | | |
|-------|---|---|---|---|---|---|---|---|---|
|       | $q^{Pw8}_{med}$ | $q^{Pw8}_{mean}$ | $q^{Pw8}_{p95}$ | $q^{Pw8}_{med}$ | $q^{Pw8}_{mean}$ | $q^{Pw8}_{p95}$ | $q^{Pw8}_{med}$ | $q^{Pw8}_{mean}$ | $q^{Pw8}_{p95}$ |
| Ours   | 1.54 | 1.74 | <u>3.11</u> | <u>1.43</u> | <u>1.80</u> | <u>3.37</u> | <u>1.27</u> | <u>1.50</u> | <u>2.40</u> |
| GNN.UP | 1.55 | 1.90 | 3.89 | 1.44 | 1.86 | 3.73 | 1.31 | 1.58 | 2.57 |
| SB.OP  | <u>1.47</u> | <u>1.72</u> | 3.18 | 1.46 | 1.87 | 3.58 | 1.32 | 1.54 | 2.54 |

**Table 7: New engine: prediction accuracy of the new predictor.**

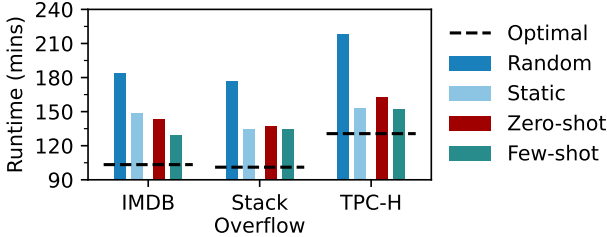

**Figure 3: Query-level routing using our proposed LCM.**

Significant reductions in the Q-error (between 20% and 40%) are observed particularly in the tail error.

*4.2.5 Results on adding a new engine.* Finally, Table 7 shows the Q-error of the LCM after including a new predictor head that represents an additional engine provisioning (*PrestoW8*) in the lakehouse. Recall that the new predictor head is only trained using 250 queries, keeping the cost of data collection low, as well as the training cost. Overall, we observe metrics for each dataset that closely match those from the few-shot experiments. This demonstrates that the learned embedding allows for including a new predictor for an engine different than those used in the LCM pre-training.

## 4.3 Effect on Query Execution Times and Routing

We analyze the LCM's effect on query-level routing both in the *zero-shot* and *few-shot* scenarios. Each query is assigned to the execution engine, which is estimated to lead to the shortest execution time.

*4.3.1 Baselines.* Two baselines are considered for query routing:
(1) *Random*: the engine is selected randomly for each query.
(2) *Static*: the complete workload is executed on the engine that minimizes the workload's total runtime. Note that this routing requires a priori knowledge in determining which engine should be used for execution.

*4.3.2 Results.* Figure 3 shows the total execution time of the considered workloads under different routing strategies. In the *zero-shot* setting, relying on the LCM for engine selection reduces total runtime by up to 25.3% over a random routing, corresponding to a 54.9 minute difference. The improved estimation accuracy in the *few-shot* scenario also translates to more accurate routing. Specifically, using a *few-shot* predictor reduces total runtime over random routing by 54.4 minutes (−29.7%) for IMDB, 41.8 minutes (−23.7%) for Stack Overflow, and 66.1 minutes (−30.4%) for TPC-H. Finally, the fine-tuned predictors also lead to similar or lower total execution time compared to static routing. For instance, on the IMDB dataset, the difference between the LCM-based and static routing grows to 18.8 minutes (12.7%) in favor of the LCM.

We remark the significant differences compared to random routing, despite the fact that queries are short-running (<1 minute) and thus limit the gains from engine selection. Scaling data sizes and raising the timeout threshold will likely offer further improvements.

## 4.4 LCM Training

Finally, Table 3 summarizes the intermediate sizes of each MLP used in the GNN and set-based model. The resulting GNN has a total of 4.7M parameters, whereas the set-based model consists of 1.6M parameters. The models are trained with the AdamW optimizer [19] and a learning rate of 0.001 for at least 200 epochs, using early stopping with a patience of 25 epochs. From each dataset considered for training, 250 queries are reserved for validation.

Using a single NVIDIA Tesla V100-SXM2 32GB GPU, the training on the complete dataset (∼17k datapoints in the training set) takes around 7 hours for the GNN and 3 hours for the set-based model. The fine-tuning process for few-shot takes ∼10min for the GNN model and ∼5min for the set-based model. Finally, the inference time on GPU for a single query is ∼4.5 ms with the GNN architecture and ∼0.7 ms with the set-based.

## 5 CONCLUSIONS

In this paper, we have presented a cross-engine optimizer for executing SQL workloads in lakehouse systems, which automates the engine selection process. This has the benefit of simplifying the lakehouse architecture and presenting it to the user as a single-endpoint interface. We have shown that combining traditional query optimization techniques with an LCM leads to enhanced prediction accuracy due to the more accurate query plan representation received and learned by the LCM. Furthermore, we proposed to formulate cost prediction across multiple engines as a multi-task learning problem, thereby avoiding the need to train engine-specific cost models and flexibly supporting the inclusion of new engine instances at a low cost.

We identify the random query generation as the main limitation, as it only provides weak control over the nature of the generated queries, both in terms of their complexity and their semantic plausibility. For example, in some cases, random predicate combinations often filter out most or even all of the data early, producing near-empty joins. In other cases, queries become full-table joins, leading to long execution times. We believe that with the increasing availability of publicly accessible real-world databases, one can design a synthetic query generator that produces a larger diversity of meaningful queries, with varying degrees of complexity. Such a query generator could then be further integrated with the cost estimator, such that the LCM's past estimation errors can guide the query generation process via reinforcement learning techniques.

In the future, we also plan to enhance the LCM with cost aspects around engine provisioning, storage, data movement, and engine load, making the routing decision more informed. Furthermore, we aim to adjust the modeling approach to estimate node-level cost and include this in an optimizer for the distributed execution of SQL queries.

# 6 APPENDIX
## 6.1 Set-based model

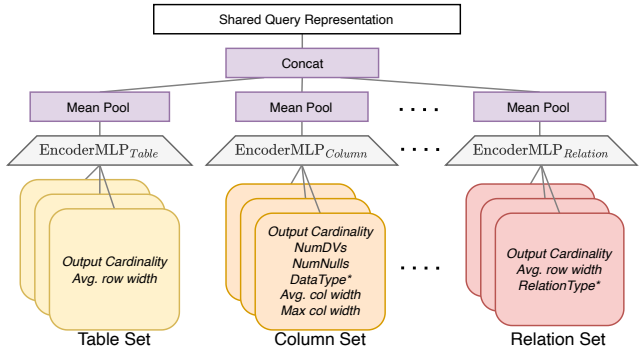

**Figure 4: Set-based approach for query representation.**

The *SB.OP* baseline is adjusted from Kipf et al. [16]. In particular, the same featurization process is employed as for the GNN architecture. However, each node belonging to the same type is treated as a set of objects, without their exact relationship being modeled. Analogous to the GNN, five sets are used for modeling, each corresponding to one of the considered node types. The resulting approach is depicted in Figure 4.

## 6.2 Results for TPC-DS and Donor

|        | TPC-DS |        |        | Donor |        |        |
|--------|--------|--------|--------|-------|--------|--------|
|        | $q_{med}$ | $q_{mean}$ | $q_{p95}$ | $q_{med}$ | $q_{mean}$ | $q_{p95}$ |
| Ours   | 1.48 | 2.48 | 5.93 | 1.59 | 1.98 | 3.94 |
| GNN.UP | 1.55 | 2.37 | 5.93 | 1.84 | 2.12 | 4.09 |
| SB.OP  | 1.45 | 2.16 | 5.25 | 1.70 | 2.00 | 3.84 |

**Table 8: Zero-shot: prediction accuracy of zero-shot models.**

|        | TPC-DS |        |        | Donor |        |        |
|--------|--------|--------|--------|-------|--------|--------|
|        | $q_{med}$ | $q_{mean}$ | $q_{p95}$ | $q_{med}$ | $q_{mean}$ | $q_{p95}$ |
| Ours   | 1.32 | 2.28 | 5.15 | 1.28 | 1.46 | 2.42 |
| GNN.UP | 1.34 | 2.01 | 4.81 | 1.35 | 1.53 | 2.57 |
| SB.OP  | 1.31 | 1.86 | 3.91 | 1.30 | 1.49 | 2.45 |

**Table 9: Few-shot: prediction accuracy of few-shot models.**

|        | TPC-DS |        |        | Donor |        |        |
|--------|--------|--------|--------|-------|--------|--------|
|        | $q_{med}^{Pw8}$ | $q_{mean}^{Pw8}$ | $q_{p95}^{Pw8}$ | $q_{med}^{Pw8}$ | $q_{mean}^{Pw8}$ | $q_{p95}^{Pw8}$ |
| Ours   | 1.42 | 2.25 | 3.57 | 1.37 | 1.59 | 2.82 |
| GNN.UP | 1.47 | 2.00 | 3.85 | 1.35 | 1.55 | 2.61 |
| SB.OP  | 1.34 | 1.77 | 2.92 | 1.38 | 1.58 | 2.66 |

**Table 10: New engine: prediction accuracy of the new predictor.**

*6.2.1 Results on LCM's accuracy.* In the zero-shot setting (see Table 8), the effect of plan optimization is similar to that observed for other datasets. The GNN using optimized plans achieves 4.6% lower $q_{med}$ for TPC-DS and 13.6% for Donor. Few-shot results (see Table 9) on TPC-DS and Donor also show patterns similar to those discussed in Sections 4.2 and 4.3. Specifically, both the LCMs' accuracy and the routings converge. For the zero-shot setup, $q_{mean}$ is reduced by 8.1% for TPC-DS and by 26.3% for Donor. Finally, Table 10 reports the accuracy of the newly included predictor head. The observed metrics closely match those for the original predictor heads.

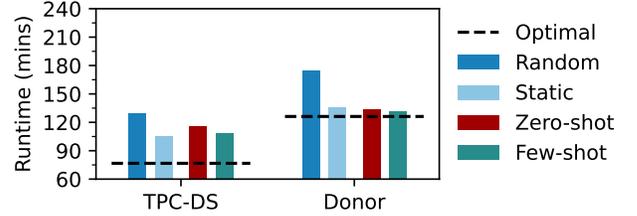

**Figure 5: Query-level routing using our proposed LCM**

*6.2.2 Effect on Query Execution Times and Routing.* Figure 5 shows the routing achieved with the proposed LCM. Using *zero-shot* predictors, the total runtime is reduced by 13.9 minutes (−10.8%) for TPC-DS and by 41.2 minutes (−23.6%) for Donor over random routing. In the *few-shot* setting, results show further improvements with 20.5 minutes (−15.9%) lower total runtime for TPC-DS and 42.6 minutes (−24.4%) for Donor. The fine-tuned predictors lead to a routing on-par with static routing.